\documentclass[letter]{aa}  
\usepackage{graphicx}
\usepackage{txfonts}
\usepackage{subcaption}
\begin{document} 
   \title{ Face changing companion of the
redback millisecond pulsar PSR J1048+2339}
   \author{Y. X. Yap\inst{1}
          \and
          K. L. Li\inst{2}
          \and
          Albert K. H. Kong\inst{1}
          \and
          J.Takata\inst{3}
          \and
          Jongsu Lee\inst{4}
          \and
          C. Y. Hui\inst{5}
          }
   \institute{Institute of Astronomy, National Tsing Hua University, Hsinchu 30013, Taiwan\\
   		\email{yapyeexuan@gapp.nthu.edu.tw}
   	\and Department of Physics, UNIST, Ulsan 44919, Korea\\
   	\and School of Physics, Huazhong University of Science and Technology, Wuhan 430074, China\\
   	\and Department of Space Science and Geology, Chungnam National University, Daejeon, Republic of Korea\\
   	\and Department of Astronomy and Space Science, Chungnam National University, Daejeon, Republic of Korea
             }
   \date{Received October 31, 2018; accepted December 20, 2018}

% \abstract{}{}{}{}{} 
% 5 {} token are mandatory
 
  \abstract{We present optical observations of the redback millisecond pulsar PSR J1048+2339, which is a 4.66 ms radio pulsar in a compact binary with an orbital period of six hours. 
%Previous sparse sampling optical observations suggest that the companion is a late-type star with a minimum mass of 0.3 \(\textup{M}_\odot\). 
 We obtained high-quality light curves of PSR J1048+2339 with the Lulin 1 m Telescope. The system shows two distinct six-hour orbital modulations, in which an ellipsoidal modulation changes into a sinusoidal-like profile in less than 14 days. In
addition to the change, the brightness of the companion increased by one magnitude, suggesting
that the latter type of modulation is caused by the pulsar wind heating of the companion and that the heating became dominant in the system. 
While the changes are not unexpected, such a timescale is the shortest among
similar systems. 
We performed modeling analysis to extract the properties of the system.
We obtained a derived pulsar mass of 2.1 \(\textup{M}_\odot\) and a companion star mass of 0.4 \(\textup{M}_\odot\) for the system.
The irradiation power increased by a factor of 6 during which the pulsar wind heating dominates.
We also report on the
two archival \textit{Chandra} X-ray observations and discuss several possibilities that might cause the
varying heating on the companion.  }
   \maketitle
%
%________________________________________________________________
\section{Introduction}

PSR J1048+2339 (hereafter J1048+2339) is a 4.66 ms millisecond pulsar that was discovered by \citet{2016ApJ...819...34C} using Arecibo Observatory. It was later confirmed as a redback system by \citet{2016ApJ...823..105D} with multiwavelength observations. 
The system has an orbital period of six hours and spin-down power $\dot{E}=1.2\times10^{34}$ erg s$^{-1}$. This object is more than just a radio pulsar; it has a gamma-ray counterpart, which is 3FGL J1048.6+2338 in the Fermi Large Area Telescope (LAT) four-year point source catalog (3FGL; \citealt{2015ApJS..218...23A}). J1048.6+2338 was originally associated with a BL Lac active galactic nuclei (AGN; NVSS J104900+233821) in 3FGL, however the improved localization in the recent preliminary LAT eight-year source list (FL8Y){\color{blue}\footnote{see https://fermi.gsfc.nasa.gov/ssc/data/access/lat/fl8y/}} ruled out the possibility that it is an AGN. In a previous optical study with the Catalina Real-Time Transient Survey (\textit{CRTS}), Palomar Transient Factory (\textit{PTF}), Sloan Digital Sky Survey (\textit{SDSS}), and Pan-STARSS survey data, the light curves indicated intrinsic variability in the system. 
While there are no significant X-ray (the Neil Gehrels \textit{Swift-XRT}) and gamma-ray (\textit{Fermi-LAT}) pulsations detected, \citet{2016ApJ...823..105D} proposed that J1048+2339 has a minimum mass of 0.3 \(\textup{M}_\odot\) and effective temperature $T_{eff}$ of 3350~K, with assumptions of a 1.4 \(\textup{M}_\odot\) mass neutron star and an inclination of 90$^{\circ}$. 
In this letter, we present the optical observations of J1048+2339 taken in 2018 March and April, in which the orbital modulation changes from ellipsoidal to pulsar wind heating in less than 14 days. We also include light curve modeling to constrain the geometry of the binary and an independent X-ray analysis based on the two archival \textit{Chandra} observations.
 
   \begin{figure*}
   \centering
   \includegraphics[width=18cm]{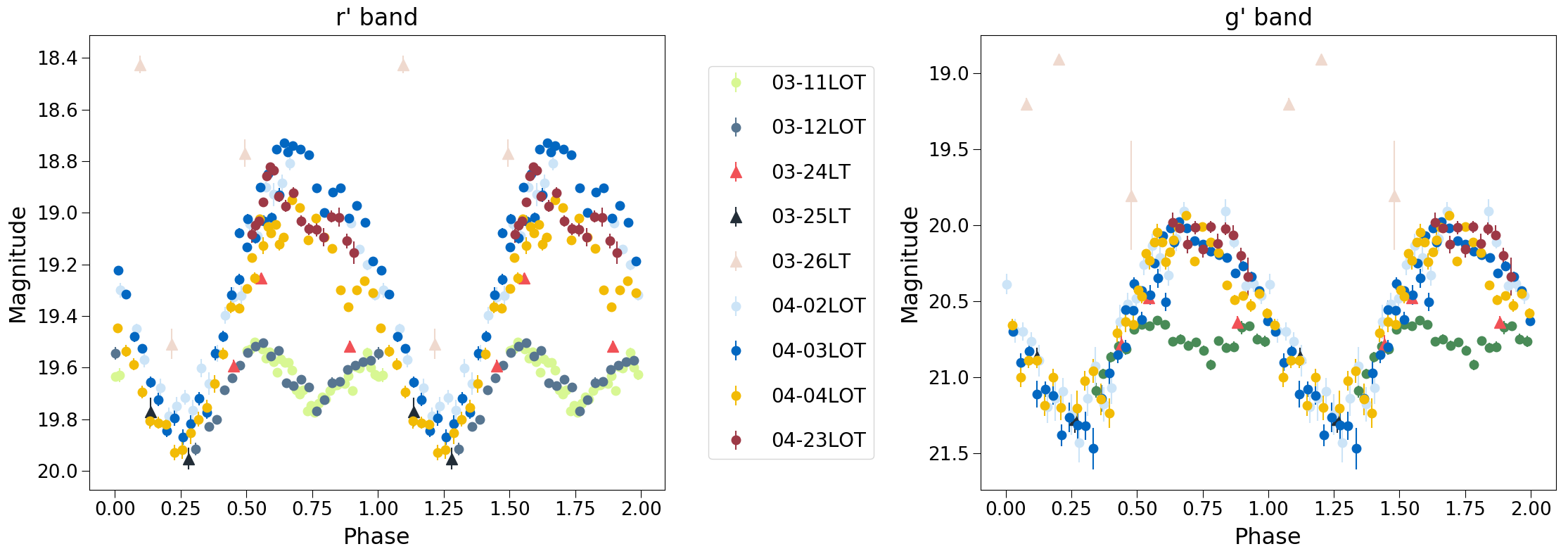}
   \caption{Light curves of PSR J1048+2339 companion star with the SDSS r$'$ and g$'$ band filter between 2018 March 11 and April 23, folded with an orbital period of 6 hours and $\textup{T}_{0}=\textup{MJD}$ 56637.598177. While the pulsar wind heating and ellipsoidal variation are clearly seen in both bands, the magnitude in the g$'$ band has larger variation compared to that of the r$'$ band. An orbital phase of 0.25 corresponds to the inferior conjunction of the companion star.          \label{fig:f1}}
    \end{figure*}

%__________________________________________________ 

\renewcommand{\arraystretch}{1.3}
   \begin{table*}
      \caption[]{Summary of observations}
         \centering
         \label{tab:1}
         \begin{tabular}{ccccccc}
                \hline\hline 
                Date & Telescope & Filter & Phase & Number of & Exposure & Apparent\\
         & & (SDSS) & range & images & (s) & magnitudes\\
                \hline
                11.3.2018 & LOT & r$'$ & 0.50-0.01 & 36 & 300& 19.50-19.77 \\
                12.3.2018 & LOT & r$'$ & 0.31-0.00 & 24 & 300& 19.50-19.92\\
        12.3.2018 & LOT & g$'$ & 0.34-0.99 & 23 & 300& 20.62-21.09 \\
        24.3.2018 & LT & r$'$ & 0.45-0.89 & 3 & 180 & 19.25-19.59\\
        24.3.2018 & LT & g$'$ & 0.44-0.88 & 3 & 180 & 20.48-20.78\\
        (25-26).3.2018 & LT & r$'$ & 0.09-0.49 & 5 & 180 & 18.43-19.95\\
        (25-26).3.2018 & LT & g$'$ & 0.08-0.48 & 5 & 300& 18.91-21.28 \\
        (2-4).4.2018 & LOT & r$'$ & 0.01-0.99 & 100 & 300 & 18.83-19.93\\
        (2-4).4.2018 & LOT & g$'$ & 0.00-0.99 & 104 & 300 & 19.91-21.80\\
        23.4.2018 & LOT & r$'$ & 0.52-0.91 & 18 & 300 & 18.82-19.16\\
        23.4.2018 & LOT & g$'$ & 0.64-0.92 & 11 & 300 & 19.98-20.34\\
                \hline 
        \end{tabular}
    \end{table*}

%

%
%______________________________________________________________

\section{Observations and data analysis}
\label{sec:obs}
\subsection{Optical observations}

We observed J1048+2339 with the Lulin 1 m telescope (LOT) on six separate nights from 2018 March 11 to April 23. Each image is exposed for 300~s and alternated between SDSS r$'$ and g$'$ filters. These raw images are then processed with IRAF V2.15 with standard calibration procedures, including bias and flat-field corrections. 
In addition, we have data from the 2 m Liverpool Telescope (LT; \citealt{2004SPIE.5489..679S}) taken in late March, which were flat- and bias-corrected. The r$'$ and g$'$ filters are alternated with 180~s exposure time, except for the g$'$ band data taken on 2018 March 25 and 26, which were exposed for 300~s per image. In total, there are 189 images (15.23 hours total integration time) with the r$'$ filter, and 147 images (12.07 hours total integration time) with g$'$ filter.  A summary of all observations is recorded in Table 1.   
We obtained the magnitude information by performing differential photometry to the reduced images with the IRAF package $phot$. 
We used ten isolated and nonsaturated stars to compute the relative magnitudes for J1048+2330. 
The relative magnitudes were then converted to apparent magnitudes using a nearby star SDSS J104840.81+234018.0 of which the magnitudes in the r$'$ band and g$'$ band are 16.53 and 17.90, respectively.
We folded the light curves using the orbital period of 0.250519045~d and ascending node $\textup{T}_{0}=\textup{MJD}$ 56637.598177 from the radio ephemeris \citep{2016ApJ...823..105D}. The folded light curves are presented in Figure \ref{fig:f1}. The quantity $\phi=0$ corresponds to the ascending node of pulsar.

\subsection{Face changing companion}
We observed a pronounced six-hour modulation in the first two nights of observation (2018 March 11 and 12). We detected two minima at $\phi=0.25$ and $\phi=0.75$, and two maxima at $\phi=0.55$ and $\phi=0$, which resemble an ellipsoidal modulation of the companion star. 
Ellipsoidal modulation is a consequence of the orbital motion for a tidally distorted star, which is commonly seen among redback systems (e.g., PSR J2129$-$0429; \citealt{2015ApJ...801L..27H}; \citealt{2016ApJ...816...74B}).  
In Figure \ref{fig:f1}, we see that the minima at $\phi=0.25$ is deeper than $\phi=0.75$ by $\sim$0.1~mag.   

\begin{figure}
\centering
\includegraphics[width=3.3in, height=2.4in]{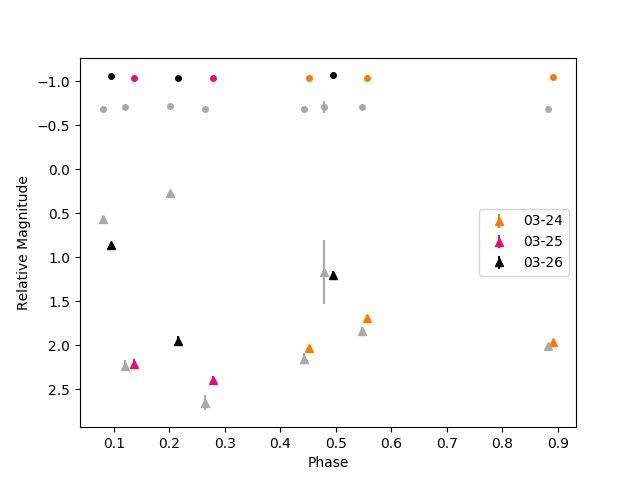} %[width=3.5in, height=2.8in]
\caption{Close examination of PSR J1048+2339 companion in late March 2018. The light curve of the companion star (triangle) and a nearby comparison star (circle) are plotted to confirm the magnitude decrease/increase at $\phi=0.21$ observed in the g$'$ (gray) and r$'$ (colored) band. The comparison star is relatively stable, indicating the observed variation at $\phi=0.21$ in J1048+2339 companion is likely genuine. }
\label{compa}
\end{figure}

Subsequent observing runs with LT were carried out from 2018 March 24 to 26. On March 24, 
we observed an increase of 0.3~mag (compared to the previous data taken at the same phase) at $\phi=0.55$ in r$'$ band. Similar changes are also observed in the g$'$ band.
On March 25, data points were taken close to the companion inferior conjunction ($\phi=0.25$), and they agree well with the minimum seen in other observations. 
On March 26, the r$'$ magnitude at $\phi=0.09$ increased by 1~mag (compared to the previous data taken at the same phase) and then decreased quickly by 1.1~mag at $\phi=0.21,$ whereas the corresponding g$'$ magnitude, which was taken slightly earlier, increased instead. 
To make sure there was no random error involved, we plot J1048+2339 against one of the comparison stars to compare their light curves (cf. Figure \ref{compa}).
The light curve of the comparison star is relatively stable, indicating the observed variation in the companion is likely genuine.
This might be because of a spontaneous flaring event of some kind of transition that arose in late March. 

J1048+2339 was observed again in early April with LOT. As shown in Figure \ref{fig:f1}, the folded light curves have a minimum at $\phi=0.25$ and a maximum at $\phi=0.65$, which are very different compared to the March light curves.
The April light curves resemble pulsar wind heating. As discussed in \citet{2016ApJ...828....7R}, a direct pulsar irradiation toward the companion star would be expected to give a peak at $\phi=0.75$, but intrabinary shock can accounted for the peak observed slightly before 0.75.  
Heating from the intrabinary shock was also discussed in  \citet{2014ApJ...785..131T} and \citet{2014ApJ...797..111L}. From our observations, the pulsar heating effect at $\phi=0.75$ was affected by the minimum in the ellipsoidal modulation. The trace of ellipsoidal modulation can still be seen even when pulsar heating becomes dominant. That implies that the flux produced from the irradiated companion takes on a considerable proportion of the total flux detected. In addition, the g$'$ band light curve has larger variation than that of the r$'$ band light curve, indicating a larger temperature variation. 
The distinct changes from March to April light curves  
suggest that a face changing mechanism took place in less than two weeks.
%This kind of flaring episode or state transition has been observed in PSR J1023+0038 (\citealt{2009Sci...324.1411A}; \citealt{2014ApJ...781L...3P}; \citealt{2014ApJ...785..131T}; \textbf{\citealt{2014MNRAS.444.1783C}}). 
The heating effect in J1048+2339 was evident up until our observation in late April.

\subsection{\textit{Chandra} observation} 

We also explored the X-ray properties of J1048+2339 
with two archival \textit{Chandra} data, which were taken on 2017-03-08 (obsID 19039; 22.6~ks) and 2017-07-04 (obsID 19038; 24.7~ks). In both observation, J1048+2339 was imaged with the back-illuminated CCD ACIS-S3. The data were reprocessed using standard \textit{Chandra} Interative Analysis of Observations (\textit{CIAO}) software and updated \textit{Chandra} Calibration Database (\textit{CALDB}). All the subsequent analysis
are restricted in 0.3-8~keV energy range. 

An X-ray point source is clearly detected at the pulsar position with no hint of any extended X-ray emission. The phase-averaged spectrum were extracted from a 3 arcsec source region. The background spectra were sampled from a nearby source-free region in each observation. After background subtraction, we obtained 91 and 122 net counts for the respective observations.
For each spectrum, we binned the data to have at least 10 counts per spectral bin. This led to an approximately Gaussian distribution of the binned data so that we can adopt chi-square as the fit statistic.
We then performed spectral analysis using the absorbed power-law model. 
We found that the best-fit spectral parameters and fluxes deduced from both observations are consistent within
the tolerance of the statistical uncertainties. Therefore, we fitted both spectra simultaneously in order to maximize 
the photon statistics and obtain tighter constraints for the spectral parameters. 

\begin{figure}
\centering
\includegraphics[width=1.9in, height=3.2in, angle=270]{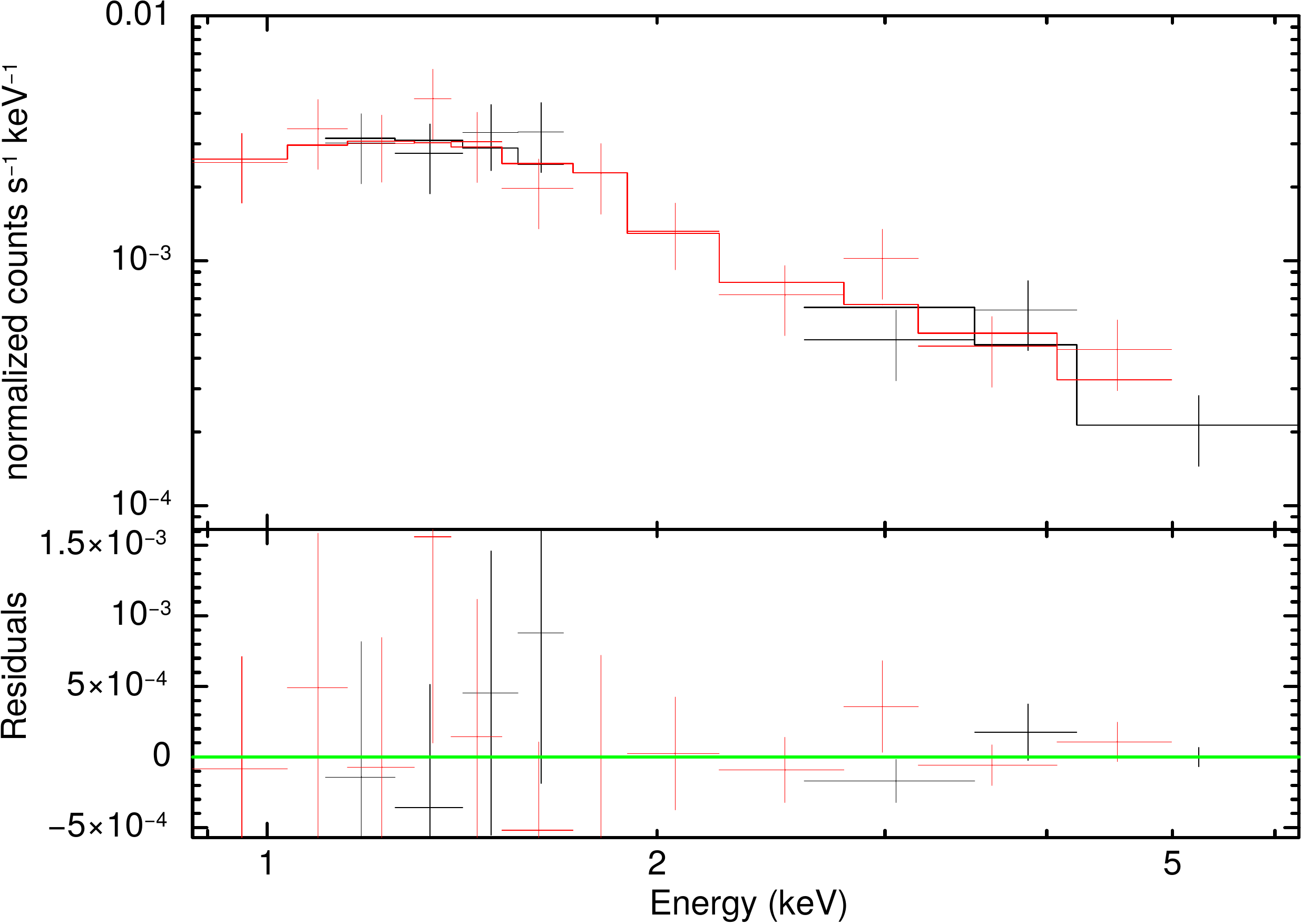}%3.2
\caption{X-ray spectra of PSR~J1048+2339 obtained from both \textit{Chandra} observations (ObsIDs: 19038 and 19039) with the best-fit 
absorbed power-law model ({\it top panel}) and the fitting residuals ({\it bottom panel}).}
\label{X_spec}
\end{figure}

As the column absorption $N_{H}$ cannot be constrained properly, we fixed it 
at the total Galactic HI column density, $N_H=2.3\times10^{20}$ cm$^{-2}$, in the direction of J1048+2339 \citep{2005A&A...440..775K}. The best-fit with a power-law yield a photon index of $\Gamma=1.64\pm0.15$ (c.f. Figure~\ref{X_spec}) and 
a goodness-of-fit of $\chi^{2}=7.64$ for 17 d.o.f. 
The photon index is similar to the typical value of redbacks \citep{2018ApJ...864...23L}. 
The unabsorbed X-ray flux in 0.3-8~keV is $f_{X}=7.25^{+0.54}_{-0.51}\times10^{-14}$ erg s$^{-1}$ cm$^{-2}$.
At a distance of 852~pc as estimated by {\it Gaia} \citep{2018AJ....156...58B}, the 0.3-8~keV X-ray luminosity is 
$L_X=6.29^{+0.47}_{-0.45}\times10^{30}$ erg s$^{-1}$. We also considered a pure thermal scenario by fitting the spectra with an absorbed blackbody model. However, the goodness-of-fit is found to be undesirable ($\chi^{2}=28.7$ for 17 d.o.f.).

\begin{figure}
\centering
\includegraphics[width=3.6in, height=3.4in]
{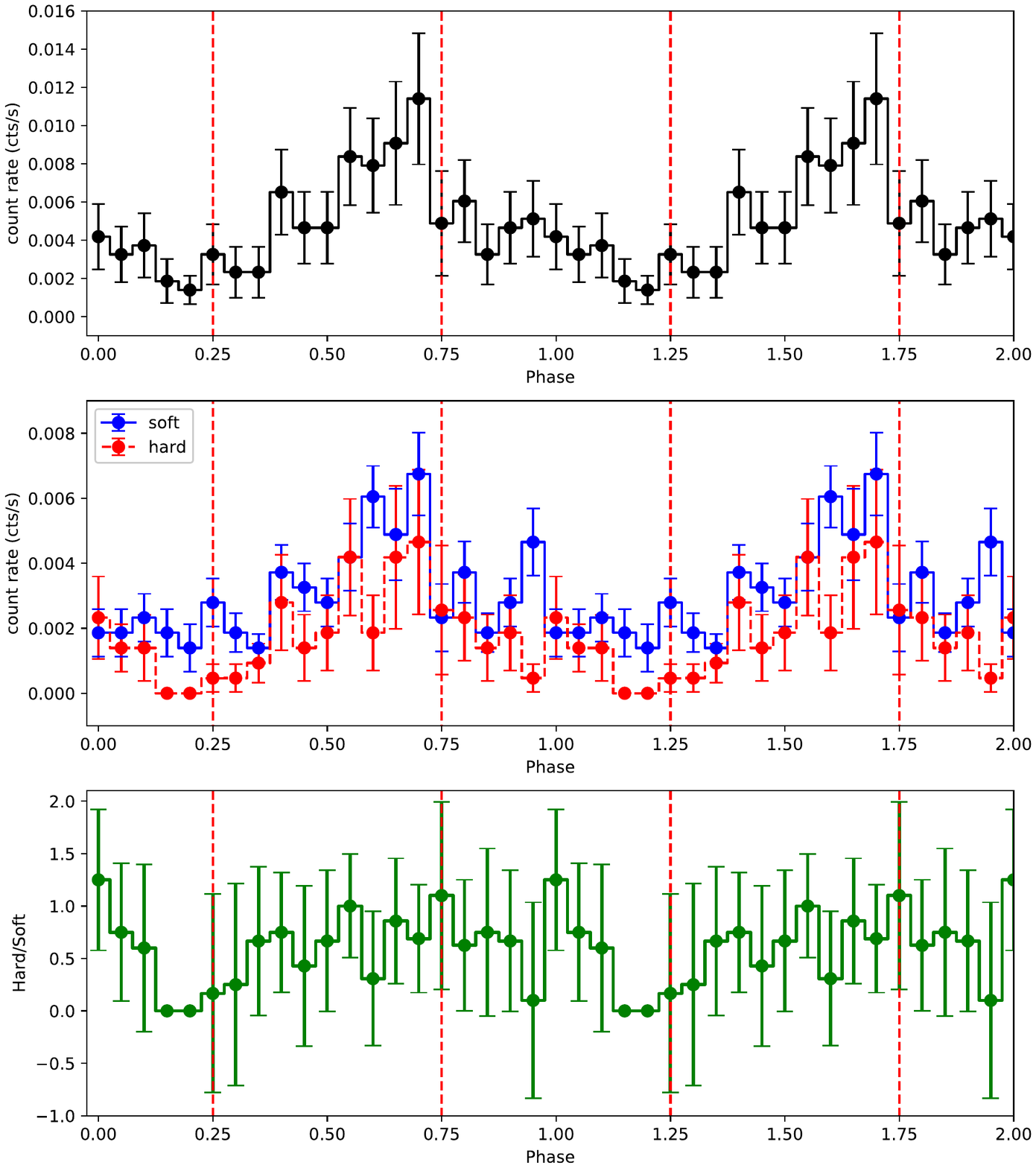} %[width=3.8in, height=4.6in]
\caption{({\it Top panel}) X-ray orbital modulation of PSR~J1048+2339 in 0.3-8~keV with both \textit{Chandra} 
observation combined. The phase zero is at same epoch as that in Figure 1. The time bin size is about $2000s$. 
The red vertical lines illustrate the locations of inferior conjunction ($\phi=0.25$) and 
superior conjunction ($\phi=0.75$) of the companion. 
({\it Middle panel}) The blue solid and red dashed lines illustrate the orbital modulation in soft band (0.3-2 keV) and hard band (2-8 keV), 
respectively. ({\it Bottom panel}) Variation of X-ray hardness across the orbit. 
}
\label{X_lc}
\end{figure}

After barycentering the arrival times in both 
observations, we folded the data from each observation at the same orbital period and $\textup{T}_{0}$ defined previously for the optical light curves.
We did not identify any significant differences between the two observations. 
In the top panel of Figure~\ref{X_lc}, we show the orbital modulation of J1048+2339 in 0.3-8~keV by combining the data from these two observations. The X-ray emission is found to attain the peak 
just before the companion enters the superior conjunction ($\phi=0.75$). We also examined if there is any differences in 
the modulation in the soft band (0.3-2~keV) and hard band (2-8~keV) (cf. middle panel of Figure~\ref{X_lc}). However, there is no 
significant variation of X-ray hardness can be found (cf. bottom panel of Figure~\ref{X_lc}). 

\section{Optical light curve modeling}
\label{sec:model}
We use the eclipsing light curve code (ELC) by \citet{2000A&A...364..265O} to generate light curve models for two groups of data according to their modulation.
We combined 2018 March 11 and 12 light curves (ellipsoidal modulation) into one group (group 1) and 2018 April 2 and 3 light curves (pulsar heating modulation) into another (group 2).
We save computing time by doing so and most importantly, this grouping is adequate for exploring the heating effect seen from the two distinct groups that are three weeks apart. 
In the modeling analysis, we assumed the following: convective envelopes for both stars, a nonspherical shape of the companion, absence of an accretion disk, and point source X-ray heating. 
Owing to the difficulty in measuring the spot locations, no spots are added to the models, although it can be accounted for the different maximum observed at the ascending and descending node of the ellipsoidal-modulated light curve. 
We also supply a SDSS atmosphere model to the ELC. 
We use the optimisers provided in the ELC to fit the folded g$'$ and r$'$ band light curves simultaneously; the parameters are listed in Table \ref{tab:2}. We first use
GridELC to search around the initial guess, by performing a “grid search” with the assigned grid size, until a minimum $\chi^{2}$ is obtained. 
We then use differential evolution Monte Carlo Markov Chain (demcmcELC) to explore the probability of these fitted parameters. 
Each fit evolved over 2000 generations with 50 members in the population.

\begin{figure*}
  \centering
  \begin{subfigure}[b]{0.475\textwidth}
    \centering
    \includegraphics[width=\textwidth]{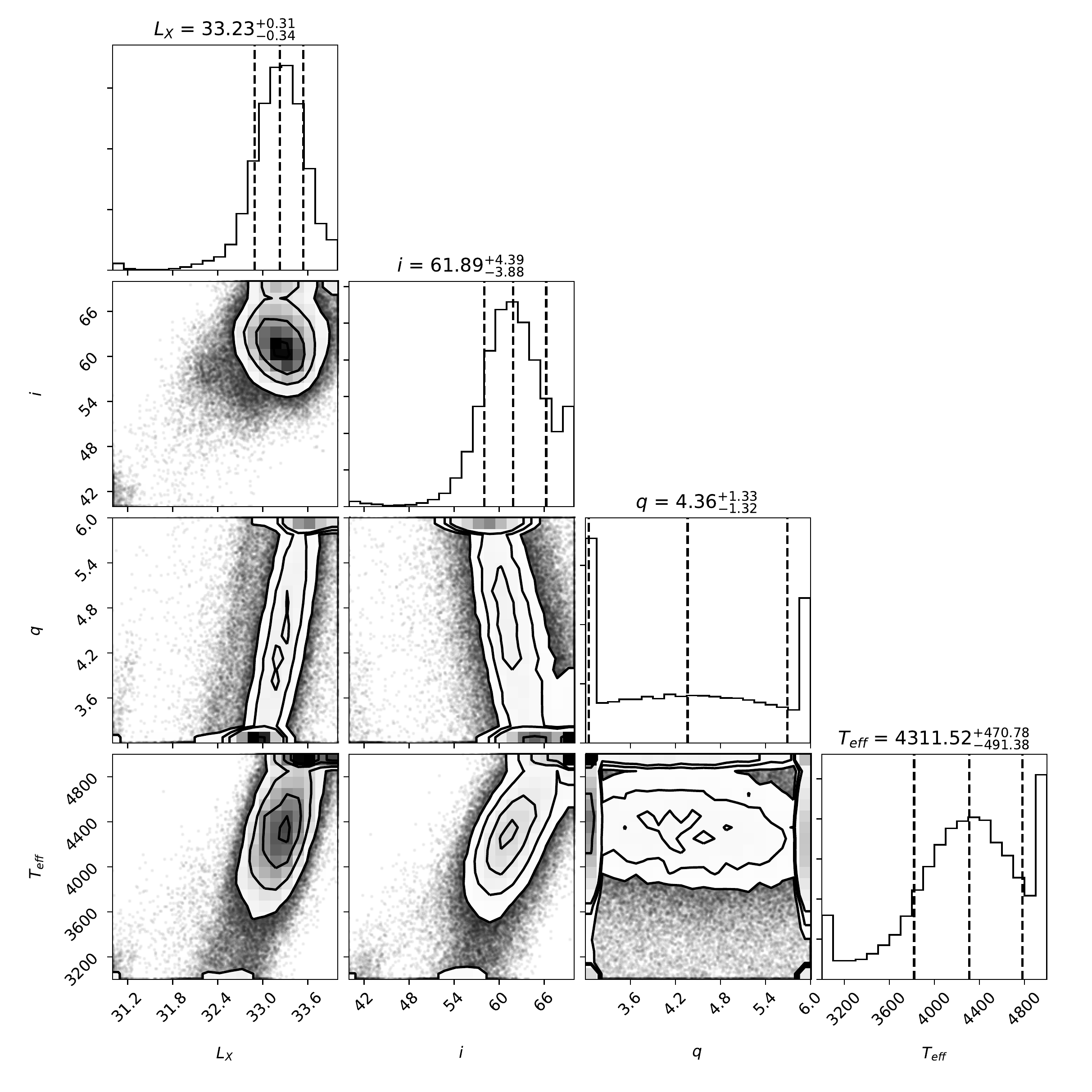}
%    \caption{Corner plot of Group 1 (ellipsoidal modulation) parameters. Roche-lobe filling factor of 1 is assumed. }%
  \end{subfigure}
  \quad
    \begin{subfigure}[b]{0.475\textwidth}
    \centering
    \includegraphics[width=\textwidth]{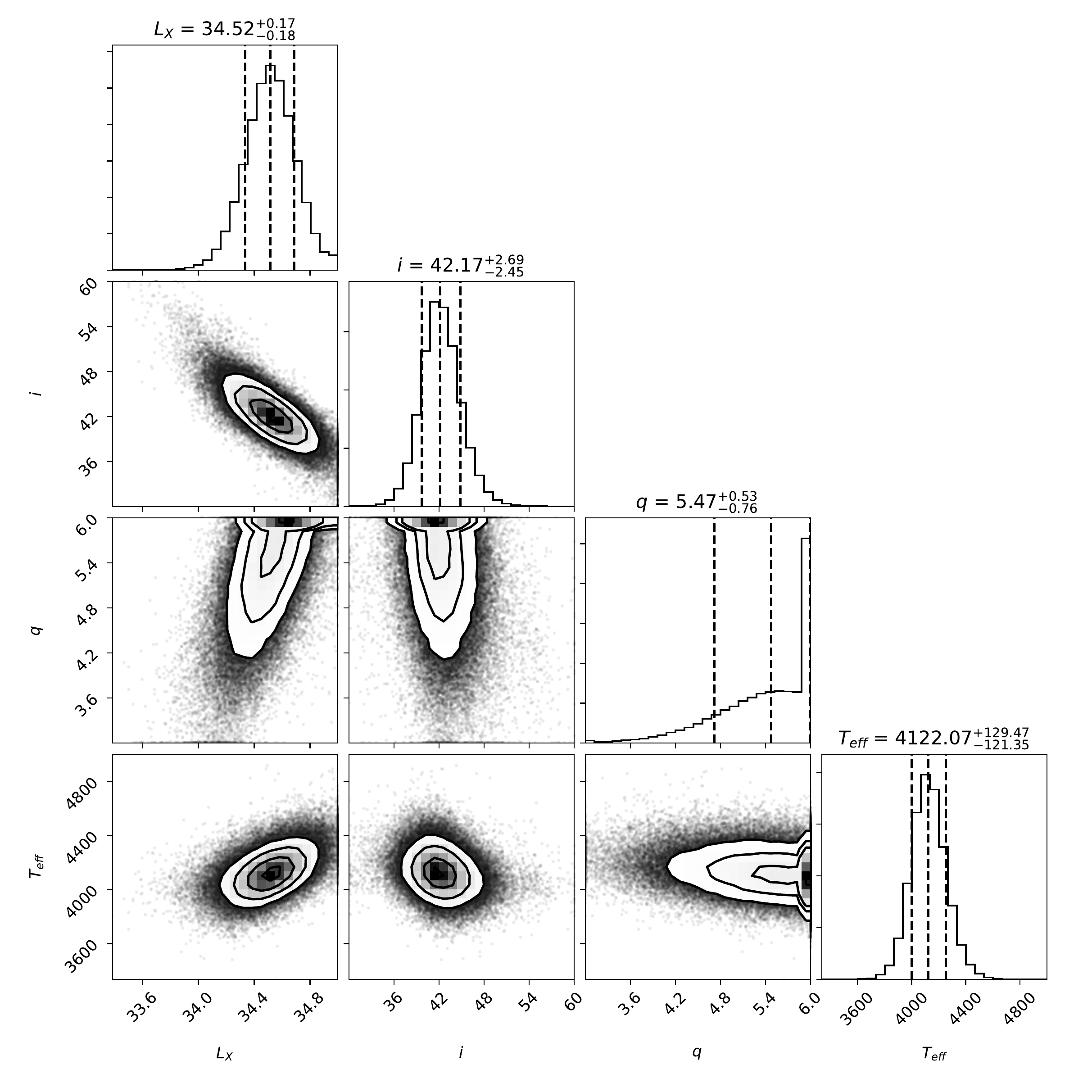}
%    \caption{Corner plot of Group 2 (pulsar heating modulation) parameters. Roche-lobe filling factor of 1 is assumed.}%
  \end{subfigure}
  \vskip\baselineskip
  \begin{subfigure}[b]{0.475\textwidth}
    \centering
    \includegraphics[width=\textwidth]{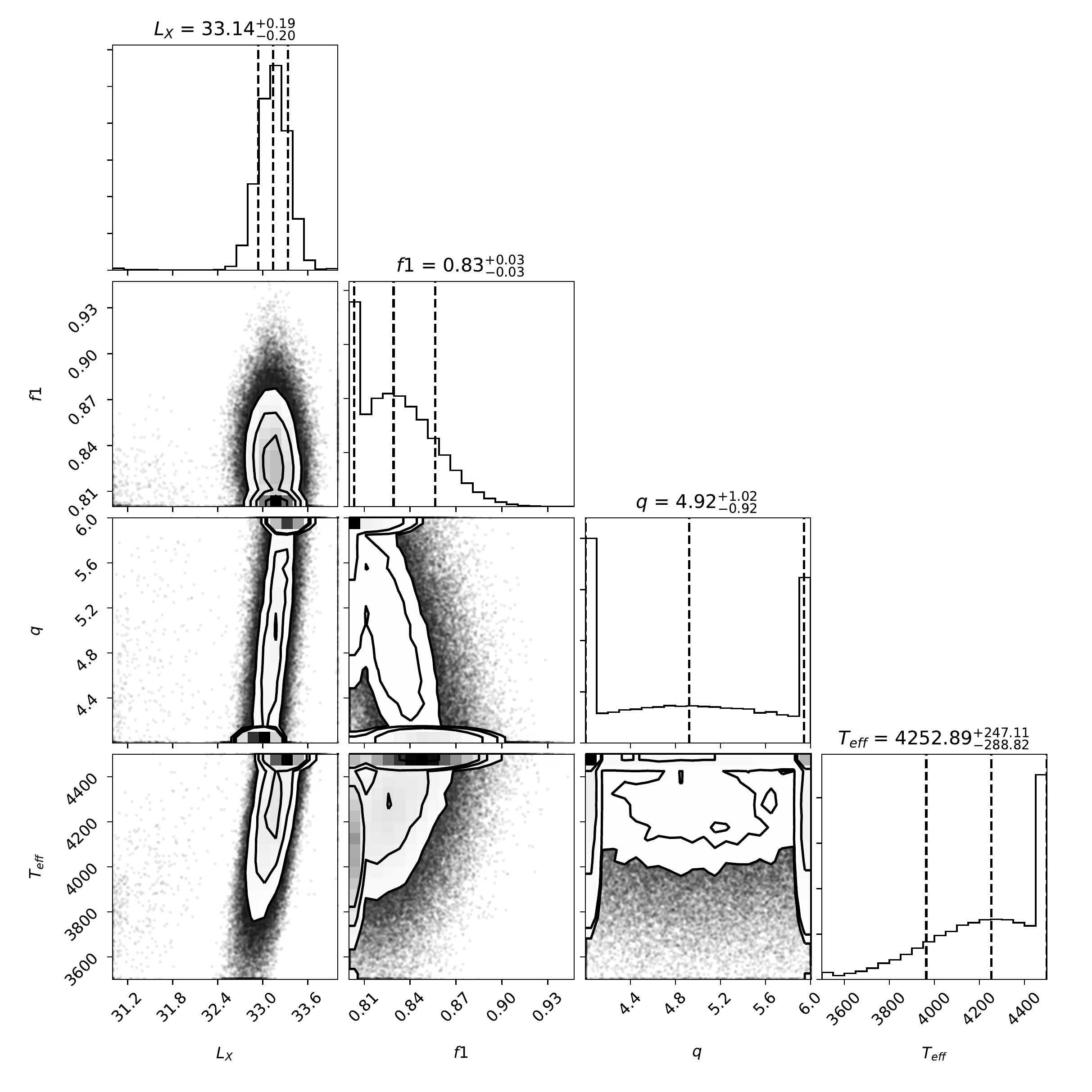}
%    \caption{Corner plot of Group 1 (ellipsoidal modulation) parameters. Inclination of 76$^{\circ}$ is assumed.}%
  \end{subfigure}
  \quad
  \begin{subfigure}[b]{0.475\textwidth}
    \centering
    \includegraphics[width=\textwidth]{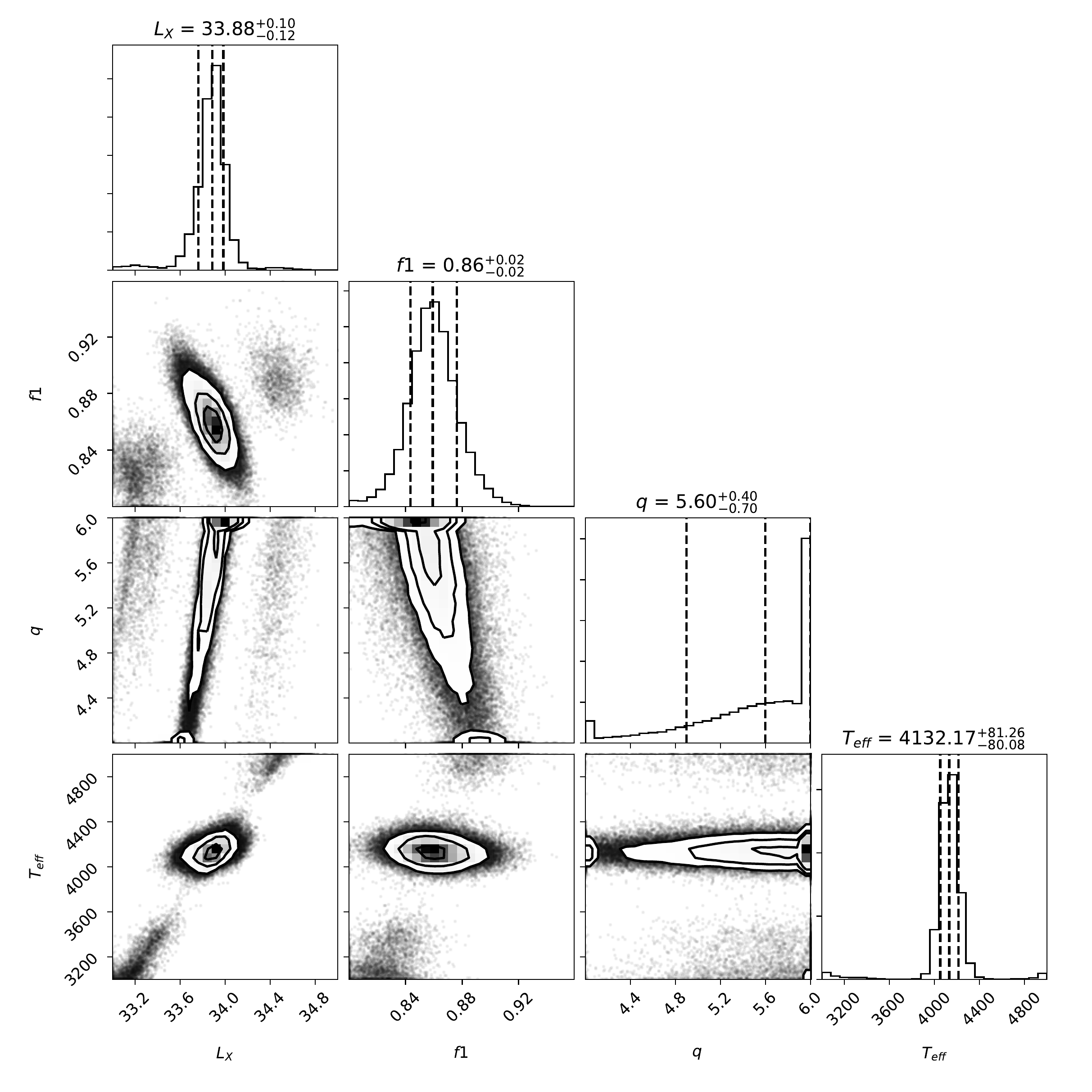}
%    \caption{Corner plot of Group 2 (pulsar heating modulation) parameters. Inclination of 76$^{\circ}$ is assumed.}%
  \end{subfigure}
  \caption{Corner plot of the fitted parameters: luminosity {\it$L_{x}$}, filling factor {\it $f_{1}$}, mass ratio {\it q}, and effective temperature {\it $T_{eff}$}. (\textit{Top left}) Roche-lobe filling factor of 1 is assumed in Group 1 (ellipsoidal modulation) fitting. (\textit{Top right}) Roche-lobe filling factor of 1 is assumed in Group 2 (pulsar heating modulation) fitting. (\textit{Bottom left}) Inclination angle of 76$^{\circ}$ is assumed in Group 1 (ellipsoidal modulation) fitting. (\textit{Bottom right}) Inclination angle of 76$^{\circ}$ is assumed in Group 2 (pulsar heating modulation) fitting. The best-fit parameter is displayed with 1$\sigma$ confidence interval for a 2D histogram. }
  \label{fig:6}
\end{figure*}

Initial fitting results inferred that the companion fully filled its Roche lobe ($f_{1}$=0.99), but other parameters were not well constrained. 
We tried fixing $f_{1}$=1 and the results are shown in Figure \ref{fig:6}(a) and (b).
For instance, the pulsar irradiation in Group 2 increased by a factor of 19, compared to that of Group 1. The fitted inclinations are 40$^{\circ}$-60$^{\circ}$ while the effective temperatures (corresponding to the night side temperature of the star) are around 4000 K. 
The mass ratio in both groups are unconstrained but both have a median around 4.4-5.5. We attribute this scenario to the lack of radial velocity measurements. At the fitted inclinations, the mass ratios imply a pulsar mass of around 3~\(\textup{M}_\odot\), which is too massive for a neutron star.
For a typical pulsar, the above mass ratios inferred a higher inclination. Furthermore, the difference in the pulsar irradiation is too large to be justified. 
In the subsequent analysis, we fixed the inclination to a theoretical upper limit of 76$^{\circ}$ for which an X-ray eclipse was not observed, using the derived pulsar mass of 2.1~\(\textup{M}_\odot\) and a companion star mass of 0.4~\(\textup{M}_\odot\), obtained from the light curve fitting. The results are shown in Figure \ref{fig:6}(c) and (d), in addition to Table \ref{tab:2}. We obtained a factor of $\sim$6 increment in the Group 2 pulsar irradiation, in comparison to the irradiation in Group 1. 
The filling factors and effective temperatures are $\sim$0.8 and $\sim$4200~K in both groups. 

%\citep{1983ApJ...268..368E} from the Roche lobe calculation

\renewcommand{\arraystretch}{1.5}

\begin{table}[!htb]
        \caption{ELC modeling results}
        \centering
        \label{tab:2}
    \resizebox{9.1cm}{!}{
        \begin{tabular}{lcc}
                \hline\hline 
                 Parameter & \tablefootmark{a}Group 1 & \tablefootmark{b}Group 2 \\
                \hline
                Pulsar irradiation (erg~s$^{-1}$) & $10^{33.14_{-0.20}^{+0.19}}$ & $10^{33.88_{-0.10}^{+0.12}}$   \\
                Roche-lobe filling factor & $0.83_{-0.03}^{+0.03}$ & $0.86_{-0.02}^{+0.02}$ \\
        Mass ratio (M$_{pulsar}$/M$_{companion}$) & $4.92_{-0.92}^{+1.02}$ & $5.60_{-0.70}^{+0.40}$ \\
        Effective temperature (K) &  $4253_{-288}^{+247}$ & $4123_{-80}^{+81}$ \\ 
                Inclination ($^{\circ}$) & fixed at 76$^{\circ}$ & fixed at 76$^{\circ}$ \\
        Derived pulsar mass &2.06 \(\textup{M}_\odot\) & 2.14 \(\textup{M}_\odot\) \\
        Derived companion mass &0.40 \(\textup{M}_\odot\)&0.41 \(\textup{M}_\odot\) \\
        \hline 
        \end{tabular}
    }
    \tablefoot{
    \tablefoottext{a}{2018 March 11 \& 12.}
    \tablefoottext{b}{2018 April 2 \& 4.}
    
    }
\end{table}

\section{Discussion}
\label{sec:discussion}

We observed a face changing companion of the redback system J1048+2339, in which the modulation changes evidently from ellipsoidal to pulsar heating. 
The changing timescale took no more than 14 days. It is not clear if the timescale of $\leqslant$14~days is only exclusive to J1048+2339. 
Based on our LT data, the unusual brightening at $\phi=0.21$ might be a precursor of a huge flaring taking place in the system.
More recently, \citet{2018ApJ...866...71C} reported an optical brightening and an X-ray flaring event of the system. The optical emission at orbital phase $\phi=0.75$ increased by 0.5 mag on 2018 April 18 (compared to their previous detection) and went back to quiescence ($ R \gtrsim 20.5~mag$) in the 2018 May 20 observation.
An X-ray flare was detected by \textit{Chandra} on 2018 July 8  at $\phi=0.9$.  By comparing with our observations, it follows that the pulsar wind heating timescale is at most $\sim$2 months.

The observed brightening of the companion star and the X-ray modulation provide us the information
of the heating source for the companion star.
From the observed and estimated properties of the system, the separation of the two stars, $a$,
and the Roche-lobe radius, $R_{rb}$, of the companion star
are estimated as $a=1.5\times~10^{11}$cm and $R_{rb}/a=0.25$.
%, respectively, where mass ratio of 5 from the ELC model is used.   
With the ELC model, we find the required irradiation luminosity of $L_{irr}\sim 10^{33-34}$~erg/s (cf. Table \ref{tab:2}),
if we assume the heating source is located at the position of the pulsar. In this case,  a fraction
of the luminosity absorbed by the companion star would be estimated by $\delta\sim (R^2_{rb}/4a^2)\sim 0.016$.
Hence, we may estimate the rate of the energy absorbed by the companion star as
$L_{ab}\sim \delta L_{irr}\sim 5\times 10^{31}{\rm erg/s}(\delta/0.016)(L_{irr}/10^{33.5}{\rm erg\cdot s^{-1}}$).

There are several possibilities  for the sources of the heating: the pulsar wind \citep{1990ApJ...358..561H}, gamma-ray radiation from  the pulsar \citep{2012ApJ...745..100T},
and the intrabinary shock emission \citep{2017ApJ...845...42S}. For J1048+2339,
the gamma-rays from the magnetosphere, for which the luminosity is usually $10\%$ of the spin-down power,
are not the main heating source, since the magnetospheric emission
are steady and cannot explain the observed brightening ($\sim$1 magnitude) of the companion star. The change of
the gamma-ray emission from the pulsar magnetosphere has been observed for young pulsar PSR J2021+4026 \citep{2017ApJ...842...53Z}. However, such a flux change is accompanied by a large glitch of the neutron star.
For millisecond pulsars, such a large glitch has not been observed yet. 
 
The observed X-ray luminosity from the intrabinary shock of this system is on the order
of $L_{X}\sim 10^{31}{\rm erg~s^{-1}}$, which is comparable to the required energy of the absorption $L_{ab}$.
Hence if the shock emission is the main heating source, the intrabinary shock should be located
near/on the companion star source. In such a case,
a natural explanation of the observed X-ray modulation with the flux peak around the superior conjunction
is a consequence of the obstruction of the X-ray emitting region
by the companion star. As demonstrated in \citet{2011ApJ...742...97B}, however, the modulation expected
by this scenario has a sinusoidal shape and the expected width of the peak ($\ge 0.5$ orbital phase)
is wider than the observed one peak (say, $\sim 0.3$). This scenario would also be difficult to explain the observed large
amplitude (factor of 10, c.f. Figure~\ref{X_lc}), unless our view is almost edge-on.
In fact, our results imply a high inclination of the system, consistent with the recent work by \citet{2018arXiv181204626S} in which a near to edge-on orbit for J1048+2339 is suggested using radial velocity measurements. 

Alternatively, the X-ray peak can be 
interpreted as the Doppler boosting effect of the pulsar wind flow \citep[see][]{2012ApJ...760...92H}.
In this scenario, the pressure of the stellar wind
is stronger than the pulsar wind pressure and the shock cone wraps the pulsar. The Doppler boosting effect
enhances (or weakens) the observed X-ray flux when the pulsar wind moves toward (or away from) the Earth.
PSR J2129-0429, for example, shows a double peak structure at a small dip at the superior
conjunction \citep{2018MNRAS.478.3987K}.
%\textbf{Double peak in the X-ray light curve of J1048+2339 is possible, however one peak is much smaller than another peak because of the asymmetry of the Doppler boosting effect.}
It has been suggested that since the spin of the companion is synchronized with the orbital period,
the stellar magnetic field of the companion star is enhanced to several kilogauss (kG) by the stellar dynamo process \citep{2017ApJ...845...42S}.
The X-ray peak (cf. Figure \ref{X_lc}) is not symmetric. Depending on the inclination angle of the binary, a double peak can be produced if the viewing angle is within the shock cone, while single peak is expected if the viewing angle is outside the shock cone.

If the intrabinary shock is located far from the companion star surface and wraps the pulsar,
the observed X-ray luminosity, $L_{X}\sim 10^{31}{\rm erg~s^{-1}}$, is insufficient to explain the heating
of the companion star. To overcome this difficulty, \citet{2017ApJ...845...42S} proposed that
a portion of pulsar wind particles duct along the magnetic field of the companion star to the
companion star surface, although it is uncertain how a fraction of the pulsar wind
particles can cross the contact discontinuity of the MHD shock \citep{2017ApJ...839...80W}.
If we assume that the pulsar wind particles carry an energy of $L_{PW}\sim L_{sd}=1.2\times
10^{34}{\rm erg~s^{-1}}$ and assume that several percent of the wind energy, which is required to explain the observed
X-ray luminosity for J1048+2239, is stopped by the intrabinary shock, a several percent of the shocked pulsar wind particles will
reach the companion star surface. The pairs trapped by the magnetic field
of the companion star will increase their pitch angle and lose their energy via synchrotron radiation as they move toward
the stellar surface.  This radiation
could be one of the sources of the heating. Since the pairs with a large pitch angle cannot reach the star surface owing to the magnetic mirror, only pairs with small pitch angle can deposit on and directly
heat up the companion surface. 

Light curve modeling shows that the heating energy increased by a factor of 6 in less than three weeks.
We speculate that this is related to the activity of the companion star, and  the magnetic field of the
companion star could play an important role by connecting to the shock region and guides the pair plasmas to the companion
star surface. For instance, after 2018 March 12, more pair plasmas are guided to the companion star. In the event that the magnetic field of the companion star is suddenly weakened, the location of the intrabinary shock moves toward the companion star surface, and increases the heating energy in late March. 
Interestingly, there is a spontaneous flaring event on 2018 March 26
before the launch of the pulsar heating modulation and the color
variation suggests that the flare emission is hot. It is likely due to
some heating effect related to magnetic activity on the tidally locked
companion instead of reprocessing. Future multicolor monitoring will
allow us to investigate the face changing mechanism of this intriguing
redback system.

Another speculative scenario is that the outflow from an unaccreting dead disk stops the pulsar wind located close to the pulsar (\citealt{2010MNRAS.406.1208D}; \citealt{2012ApJ...745..100T}). In this scenario, the irradiated gamma rays on the disk are absorbed  by the disk, if the column density is greater than a critical value  of $\sigma\sim 60 h/l~{\rm g~cm^{-2}}$, where $h$ and $l$ is the thickness of the disk and propagating length of the gamma rays in the disk. The absorbed energy will be converted into the outflow from the disk, which could block the pulsar wind near the pulsar. After 2018 March 12, the disk column density could be lower than the critical value, and no formation of the outflow from the disk, therefore allowing the intrabinary shock to move toward the companion star and thus increase the heating.

\begin{acknowledgements}
      We thank K. S. Cheng for helpful discussions. This work has made use of data collected at Lulin Observatory, partly supported by the Ministry of Science and Technology of the Republic of China (Taiwan) through grant 105-2112-M-008-024-MY3. The scientific results reported in this article are based on data obtained from the Chandra Data Archive. The Liverpool Telescope is operated on the island of La Palma by Liverpool John Moores University in the Spanish Observatorio del Roque de los Muchachos of the Instituto de Astrofisica de Canarias with financial support from the UK Science and Technology Facilities Council. Y.X.Y and A.K.H.K. are supported by the Ministry of Science and Technology of the Republic of China (Taiwan) through grants 105-2119-M-007-028-MY3 and 106-2628-M-007-005. J.T. is supported by NSFC grants of Chinese Government under 11573010,  1166116\ 1010, U1631103, and U1838102. J.L. is supported by BK21 plus Chungnam National University and the National Research Foundation of Korea grant 2016R1A5A1013277. C.Y.H. and K.L.L are supported by the National Research Foundation of Korea grant 2016R1A5A1013277.
    
\end{acknowledgements}

\bibliographystyle{aa} 
\bibliography{ref}

\clearpage

\end{document}